\documentclass[useAMS,usenatbib]{mn2e}

\usepackage{graphicx}	
\usepackage{amsmath}	
\usepackage{amssymb}	

\usepackage{graphicx}	

\defcitealias{1989A&A...226..366R}{RR05}

\title[Asymmetries in the Aldebaran photosphere]{Evidence of asymmetries in the Aldebaran photosphere from multi-wavelength lunar occultations}

\author[A. Richichi et al.]{
A. Richichi,$^{1}$\thanks{E-mail: andrea4work@gmail.com}
V. Dyachenko,$^{2}$
A.K. Pandey,$^{3}$
S. Sharma,$^{3}$
O. Tasuya,$^{1}$
Y. Balega,$^{2}$
\newauthor
A. Beskakotov$^{2}$
D. Rastegaev$^{2}$
and V.S. Dhillon$^{4,5}$
\\
$^{1}$National Astronomical Research Institute of Thailand,
191 Huay Kaew Road,
Chiang Mai 50200 Thailand\\
$^{2}$Special Astrophysical Observatory, 
Nizhnij Arkhyz, Karachai-Cherkessian Republic,
Russia 369167 \\
$^{3}$Aryabhatta Research Institute of Observational Sciences,
Manora Peak, Naini Tal, 263002 India\\
$^{4}$ Department of Physics and Astronomy, University of Sheffield, Sheffield S3 7RH, UK\\
$^{5}$ Instituto de Astrof{\' i}sica de Canarias, E-38205 La Laguna, Tenerife, Spain
}

\date{Accepted XXX. Received YYY; in original form ZZZ}

\pubyear{2016}

\begin{document}
\label{firstpage}
\pagerange{\pageref{firstpage}--\pageref{lastpage}}
\maketitle

\begin{abstract}
We have recorded three lunar occultations of Aldebaran ($\alpha$~Tau) at
 different telescopes and using various band-passes, from the ultraviolet to
  the far red. The data have been analyzed using both model-dependent 
  and model-independent methods. 
  The derived uniform-disc angular diameter values have been converted 
  to limb-darkened values using model atmosphere relations, and are found 
  in broad agreement among themselves and with previous literature values. The limb-darkened diameter is about 20.3~milliarcseconds 
  on average . 
  However, we have found indications that the photospheric brightness profile of 
  Aldebaran may have not been symmetric, a finding already reported
  by other authors for this and for similar late-type stars.
  At the sampling scale of our brightness profile, 
  between one and two milliarcsecond, 
  the uniform and limb-darkened disc models may not
  be a good description for Aldebaran. The asymmetries appear to
  differ with wavelength and over the 137 days time span of our
  measurements.
  Surface spots appear as a likely explanation for the differences 
  between observations and the models.
\end{abstract}

\begin{keywords}
occultations -- stars: individual: Aldebaran -- stars: atmospheres
\end{keywords}



\section{Introduction}
Aldebaran ($\alpha$~Tau) is one of the brightest and most distinctive 
stars in the sky, and as a result it has one of the longest records 
of observations and publications. Being a K5 giant star and located at
just 20\,pc from the Sun, it has also one of the largest angular
diameters among all stars and it has therefore been the subject
of numerous measurements in this sense using a number of techniques.
Since Aldebaran is located on the Zodiac, it is regularly occulted
by the Moon. We are at present in the middle of one such series
of occultations, which will last until early 2018.

\citet[RR05 hereafter]{2005A&A...433..305R} presented accurate
lunar occultation (LO) and long-baseline interferometry measurements obtained in the near-infrared, and discussed them in the context of previously available determinations. They concluded that the limb-darkened diameter of
$\alpha$~Tau is $20.58\pm0.03$ milliarcseconds (mas), or 44~$R_\odot$.
Photometric variability is less than 0.01\,mag and the diameter is 
assumed to be reliably constant. Differences in the angular diameter
values available in the literature are indeed present and sometimes
significant when taken at face value, however they can often be
justified in terms of uniform disc to limb-darkening corrections or
by intrinsic limitations in the accuracy.

We have recorded three LO events in the present series, and we
present here their detailed analysis. Our aim is not so much to
confirm or refine the angular diameter determination, but rather
to investigate possible asymmetries or surface structure features
in the photosphere of this giant star. Indications in this sense
had already been presented by \citetalias{1989A&A...226..366R}.

\section{Observations and Data Analysis}\label{sect:obs_data}
We recorded three LO light curves of $\alpha$~Tau:  in October 2015
 using the SAO 6-m telescope in Russia,  and in March 2016 using
the Devasthal 1.3-m telescope in India and the TNT 2.4-m telescope 
in Thailand. Details of the observations are provided in 
Table~\ref{tab:observations}.

\begin{table*}
	\centering
	\caption{Observations log}
	\label{tab:observations}
	\begin{tabular}{llll} 
\hline
Telescope & SAO 6-m & 1.3-m & TNT 2.4-m \\
Site & Nizhny Arkhyz, Russia & Devasthal, India & Doi Inthanon, Thailand \\
\hline
Coordinates & 41$\degr$26$\arcmin$E, 43$\degr$39$\arcmin$N, 2100m &
79$\degr$41$\arcmin$E, 29$\degr$22$\arcmin$N, 2450m & 
98$\degr$29$\arcmin$E, 18$\degr$34$\arcmin$N, 2450m \\
Date, Time (UT) & 2015-10-29, 23:32:01 & 2016-03-14, 14:40:29 & 2016-03-14, 15:15:32 \\
Event  & Reappearance & Disappearance & Disappearance \\
Predicted PA, Rate & 229$\degr$,$-$0.616 m/ms & 99$\degr$, 0.732 m/ms & 125$\degr$, 0.734 m/ms \\
Detector & ANDOR & ANDOR & ULTRASPEC \\
Filter ($\lambda$, $\Delta\lambda$) (nm) & Filterless & R (640, 130) & 
$u\arcmin$ (356, 60) \\
$\lambda_{\rm eff}$ (nm) & 752 & 644 & 371 \\
Sampling (ms) & 2.58 & 1.85 & 6.29 \\
CAL, DIF Sampling (mas)  & 0.90 & 0.76 & 2.57 \\
\hline
\end{tabular}
\end{table*}

At SAO, a commercial 512x512 pixels Andor iXon Ultra DU-897-CS0 detector was used. For this 
observation we used binning 16x16, readout rate 17~MHz, electron multiplying gain = 100, shift 
speed = 0.5~$\mu$s, 0.1~ms exposure time, kinetic regime. This resulted in a kinetic cycle time  of 
2.58~ms. The readout noise at this rate was 93~e$^{-}$.The result was a FITS data cube, with 32x32 pixel 
size and 150,000 frames.
At Devasthal, a similar detector was used, namely a  512x512 pixels frame transfer ANDOR iXon 
EMCCD (DU-897E-CS0-UVB-9DW). For this observation, the central 32x32 pixels were used in 2x2 
binning mode. Readout rate 10~MHz, shift speed 0.9~$\mu$s, 1.38~ms exposure time resulted in a 
kinetic cycle time  of 1.85~ms. The final image was a FITS data cube containing 45,000 frames.
At TNT, we used the ULTRASPEC frame-transfer EMCCD imager 
\citep{2014MNRAS.444.4009D} in the so-called 
drift mode already used previously for LO 
\citep{{2014AJ....148..100R},{2016AJ....151...10R}}.
 We used a window of 8x8 pixels 
($3\farcs6$ x $3\farcs6$ on the sky), and a SDSS $u\arcmin$ filter. The resulting FITS data cube consisted of 9371 frames, 
with sampling time of 6.288~ms and integration time of 6.123~ms.

In all three cases, we built an effective wavelength response of the instrument by convolving the 
filter with the CCD quantum efficiency and  the optics. The effective (weighted average) wavelengths $\lambda_{\rm eff}$  for 
each case are listed in Table~\ref{tab:observations}. We also included in our analysis the effects of the finite integration 
time, and of the primary diameter and obstruction.

The data cubes were trimmed to include only a few seconds around the occultation event, and 
converted to light curves using a mask extraction tailored to the seeing and image motion, as 
described in \citet{2008A&A...489.1399R}. 
The light curves were then analyzed using several methods. Firstly, 
a least-square model-dependent (LSM) analysis was used, the details of which are given in \citet{1992A&A...265..535R}. 
This approach uses a uniform-disc (UD) model of the stellar disc with its angular diameter as 
a free parameter, and achieves convergence in $\chi^2$ based
a noise model built from data before and after the occultation.
Among other parameters, this method allows to determine in principle
also the actual slope of the lunar limb from the comparison between
predicted and fitted lunar rate. 
However, in the specific case of a large angular
diameter such as that of $\alpha$~Tau all diffraction fringes except at the
most the first one (see below) are almost completely erased, and
this benefit of the LSM method cannot be realized (see
also Sect.~\ref{sect:ud-ld}).
Additionally, although our implementation of the LSM method allows
in principle  to
partly account for scintillation by means of interpolation by Legendre polynomials, for the same reason as above this cannot be
done in practice for the $\alpha$~Tau LO data.

Secondly, we used a composite algorithm (CAL) which provides a model-independent brightness 
profile of the source in the maximum-likelihood sense 
\citep{1989A&A...226..366R}. Finally, we also used a simple 
differentiation (DIF) to reconstruct the brightness profile in a model-independent fashion. This latter 
method is applicable when the occultation can be described by simple geometrical optics. This is the 
case when the source angular diameter $\phi$, the wavelength $\lambda$ and the distance to the Moon D satisfy 
the relation $\phi > \sqrt{\lambda/{\rm D}}$. With an angular diameter of about 20~mas, Aldebaran satisfies this 
relationship marginally in the far red, and completely in the ultraviolet. The difference between the 
CAL and the DIF methods is that the first one modifies an initial brightness profiles with small steps 
during a large number of iterations (typically thousands), resulting in a relatively smooth profile; the 
second method, instead, performs a single differentiation operation but is affected by  point-to-point noise in the data. 
This noise can be reduced by rebinning the data before differentiation, 
at the 
expense however of the final angular resolution.

With the 
LSM method, the achieved angular resolution is related to the time sampling but also to the quality 
of the fit (expressed by the normalized   $\chi^2$ ) and the signal-to-noise ratio  (SNR) of the data. In practice, this 
is the method which will yield the best resolution and accuracy, at least formally. For CAL and DIF, 
the resulting brightness profiles  have a step in angular resolution which is related to the 
sampling time, to the apparent speed of motion of the lunar limb, and to the distance to the Moon. 
The theoretical angular sampling of the brightness profiles for these two methods are listed in Table~\ref{tab:observations}.
In case of data rebinning, the angular step of the CAL and DIF profiles will be reduced accordingly.

\section{Results}\label{sect:results}

\begin{figure}
	\includegraphics[width=\columnwidth]{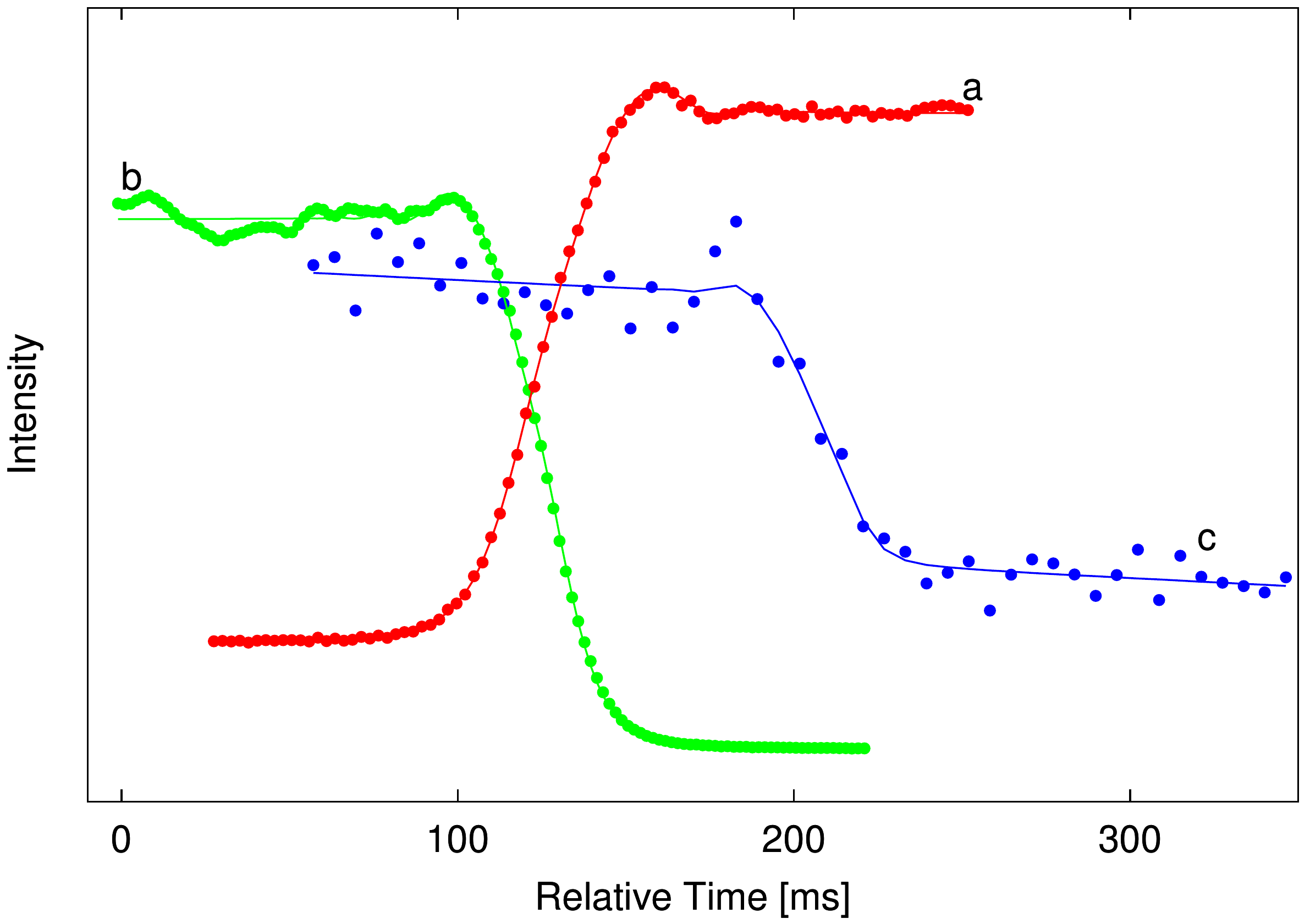}
    \caption{Light curves (points) marked as a, b, c, from the SAO 6-m, the Devasthal 1.3-m and the TNT 2.4-m telescopes, 
respectively. The data have been shifted in time, scaled and shifted in intensity, to fit in a single figure. The solid 
lines are the best fit by a uniform-disc (UD) model in each case, as discussed in the text.}
    \label{fig:figure1}
\end{figure}

The three data sets are shown in Fig.~\ref{fig:figure1}, rescaled and offset by arbitrary amounts to fit in a single, compact figure. 
Some aspects of the data can be appreciated prior to any quantitative analysis, such as for example 
the progressive transition from the diffraction to the geometrical regime. The SAO 6-m curve has a 
redder effective wavelength than the Devasthal 1.3-m curve, and indeed it shows a slightly more 
pronounced first fringe. It is however surprising that the TNT 2.4-m curve, which should be 
completely within the geometrical optics regime, appears to show in fact a  diffraction fringe.
It is also evident how the telescope diameter strongly affects the level of scintillation, 
as expected: with comparable wavelengths, the
SAO 6-m data have indeed significantly 
lower scintillation than those from the Devasthal 1.3-m.
In the remainder of this section, we report and illustrate the results of the quantitative data analysis 
of these curves, first by the model-dependent and then by the model-independent methods.

\subsection{Uniform- and Limb-Darkened Disc Diameter}\label{sect:ud-ld}
We report on the data analysis using the LSM method first, the natural outcome of which is the best 
fitting angular diameter on the basis of a noise model built individually for each light curve and each 
instrument as detailed above. We assume a stellar model with a uniform disc (UD), since this is easily 
described by one parameter only, i.e. the angular size. In reality 
the star 
is generally better described in terms of a limb-darkened disc (LD). Although it is possible to describe 
LD brightness profiles analytically, this requires a number of additional parameters and in practice 
their effect on the fitting process is not sufficient to obtain 
well-determined values unless the light 
curve has a very high SNR. We thus follow the customary approach of determining UD diameters 
first. The results are listed in Table~\ref{tab:UD_results}.
We note that, due to the almost complete suppression of the
diffraction fringes in the case of $\alpha$~Tau, the 
actual rate of the event is not a completely independent parameter
as in other LO light curves of sources with smaller angular
diameters. In fact, in this particular situation the rate is 
correlated 
to the angular size. For this reason, we have decided to keep
the rate to its predicted value in the three fits.  More
comments about this are given in Sect.~\ref{sect:discuss}.

\begin{table}
	\centering
	\caption{Uniform and limb-darkened diameter values}
	\label{tab:UD_results}
	\begin{tabular}{lccc} 
\hline
Telescope & SAO 6-m & 1.3-m & TNT 2.4-m \\
\hline
$\lambda_{\rm eff}$ (nm) & 752 & 644 & 371 \\
UD (mas) & 19.12$\pm$0.02 & 18.40$\pm$0.04 & 17.78$\pm$0.38 \\
SNR & 165.5 & 65.3 & 11.4 \\
Normalized $\chi^2$   & 1.21 & 1.76 & 0.94 \\
(LD/UD)$_{\rm eff}$ (see text) & 1.07 & 1.09 & 1.15 \\
LD (mas) & 20.42$\pm$0.02 & 20.06$\pm$0.04 & 20.48$\pm$0.44 \\
\hline
\end{tabular}
\end{table}

UD diameters are wavelength dependent, and as expected the three values listed in Table~\ref{tab:UD_results} do not 
agree with each other given the different band-passes. It can be however appreciated how the UD 
diameter value increases monotonically with $\lambda_{\rm eff}$. 
It can also be noted that the accuracy of the 
resulting diameter value is closely related to the SNR, as discussed 
in Sect.~\ref{sect:obs_data}. In order to compare 
results obtained at various wavelengths and to test atmospheric models, it is useful to convert the 
wavelength-dependent UD values to their limb-darkened (LD) diameter equivalent. 

It is 
customary to generate LD/UD coefficients from model atmospheres. 
\citet{2000MNRAS.318..387D} derived 
detailed LD/UD coefficients as a function of wavelength for a large grid of stellar atmospheric 
models, based on the atlas distributed by R.L. Kurucz on CD-ROMs in 1993. The coefficients are 
tabulated in discrete steps of effective temperature T$_{\rm eff}$, 
of surface gravity log~$g$, and of metallicity 
Z=log [Fe/H].  
For the temperature, we adopt the value T$_{\rm eff}$=3920$\pm$15~K by
\citet{1991A&A...245..567B}, 
which is in excellent agreement with the value of 3934$\pm$41~K derived 
 by \citetalias{1989A&A...226..366R}.
For  log~$g$, we adopt  the value of  1.25$\pm$0.49 by \citet{1993MNRAS.264..319B}. There 
are numerous other references for these quantities, but they generally agree within the errors with 
our adopted values. The metallicity of $\alpha$~Tau is quoted in the literature with a range of values 
ranging from solar to less than half solar 
\citep{1992A&AS...95..273C}. 
Recently, \citet{2014A&A...564A.133J}
have analyzed the data available on the metallicity of a number of 
GAIA benchmark stars 
using several different methods. They find Z=$-0.37\pm0.02$ for Aldebaran, and this is our adopted value 
as well. 

\begin{figure}
	\includegraphics[width=\columnwidth]{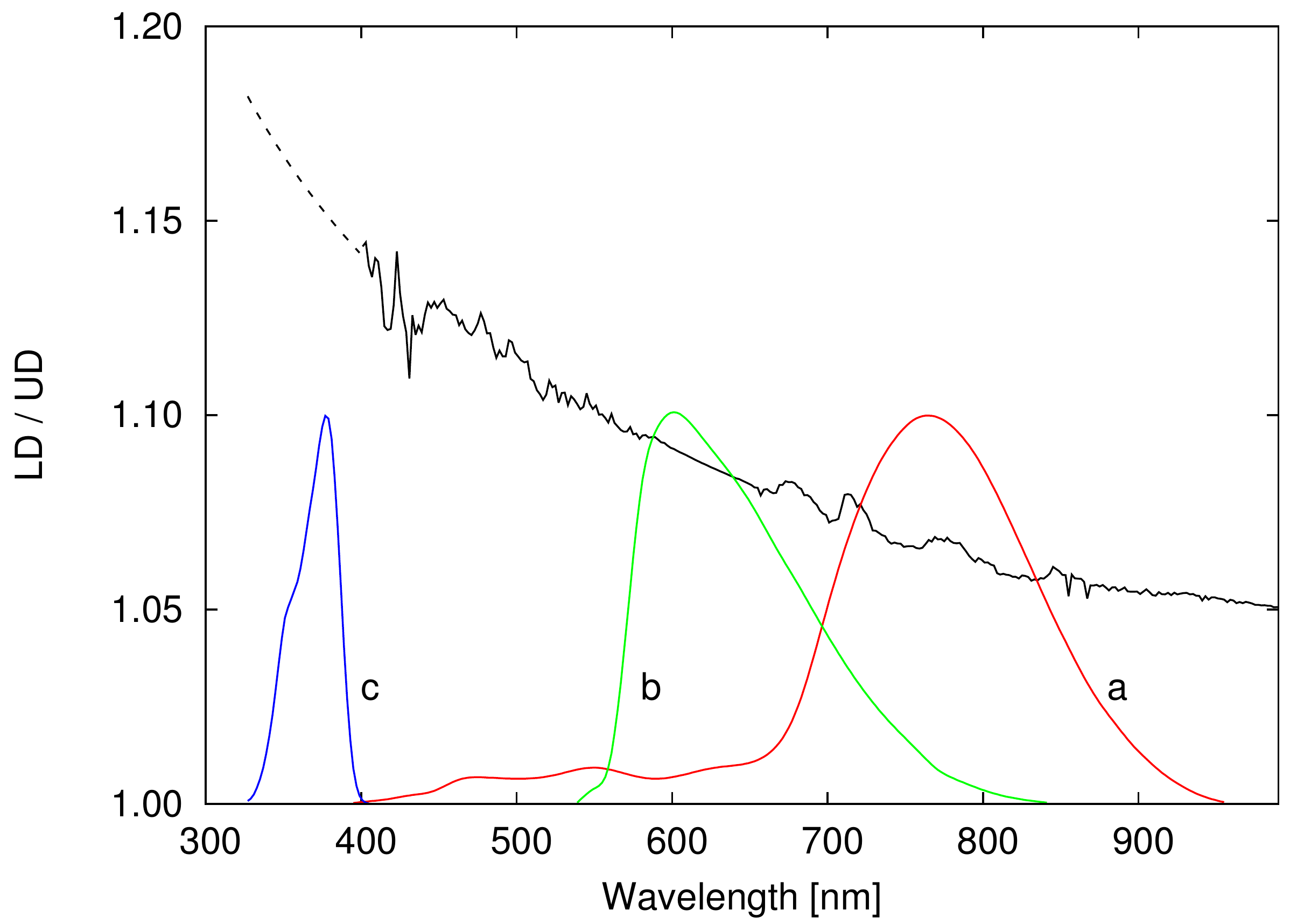}
    \caption{Solid curve: the monochromatic LD/UD coefficients, interpolated from the database of Davis et al (2005) for 
our assumed Aldebaran atmospheric parameters. Dashed curve:  extrapolation
 below 400~nm. The three 
curves labeled a, b, c are shown in arbitrary vertical units: 
they represent the total transmission of the 
instruments at SAO, Devasthal, and TNT, respectively. 
They include the effects of CCD, optics and filter.}
    \label{fig:figure2}
\end{figure}

We thus selected those curves among those provided by 
\citet{2000MNRAS.318..387D}  which bracket our 
adopted T$_{\rm eff}$, log~$g$ and Z, and proceeded to interpolate between them. The resulting LD/UD 
correction as a function of wavelength is shown in  Fig.\ref{fig:figure2}. It can be noted that the \citet{2000MNRAS.318..387D} 
data stop at $\lambda$=400~nm. Since our TNT observation extends to about 
$\lambda$=325~nm, we took the simple 
approach of extrapolating the curve, as also shown in Fig.\ref{fig:figure2}.

We note that 
as long as the LD/UD uncertainty is less than 0.03,
which seems a reasonable upper limit also in the
case of the ultraviolet extrapolation,
the effect on the error in the LD diameter is less than
the last digit shown in Table~\ref{tab:UD_results}.

\subsection{Applicability and Discrepancies of Disc Models}\label{sect:ld-discrep}
It must be remarked that the LD/UD coefficients are strictly applicable only for monochromatic 
wavelengths, while our data and thus our UD diameter results are for broad-band filters. To account 
for this, we have computed an effective LD/UD correction for each of the three LO cases, by 
weighing the LD/UD curve by the effective transmission of filter, CCD and optics, and normalizing. 
The results are listed in Table~\ref{tab:UD_results} as (LD/UD)$_{\rm eff}$. Using these corrections, we derive LD diameter values 
which are also listed in Table~\ref{tab:UD_results}. It can be
seen that the LD values from SAO and TNT are consistent
among themselves and in some agreement with 
the LD diameter of $20.58\pm0.03$~mas reported by 
\citetalias{1989A&A...226..366R}
- although not within the errors in the case of the SAO result. 
The result from the Devasthal 1.3-m telescope however has to
be considered discrepant.

\begin{figure}
	\includegraphics[width=\columnwidth]{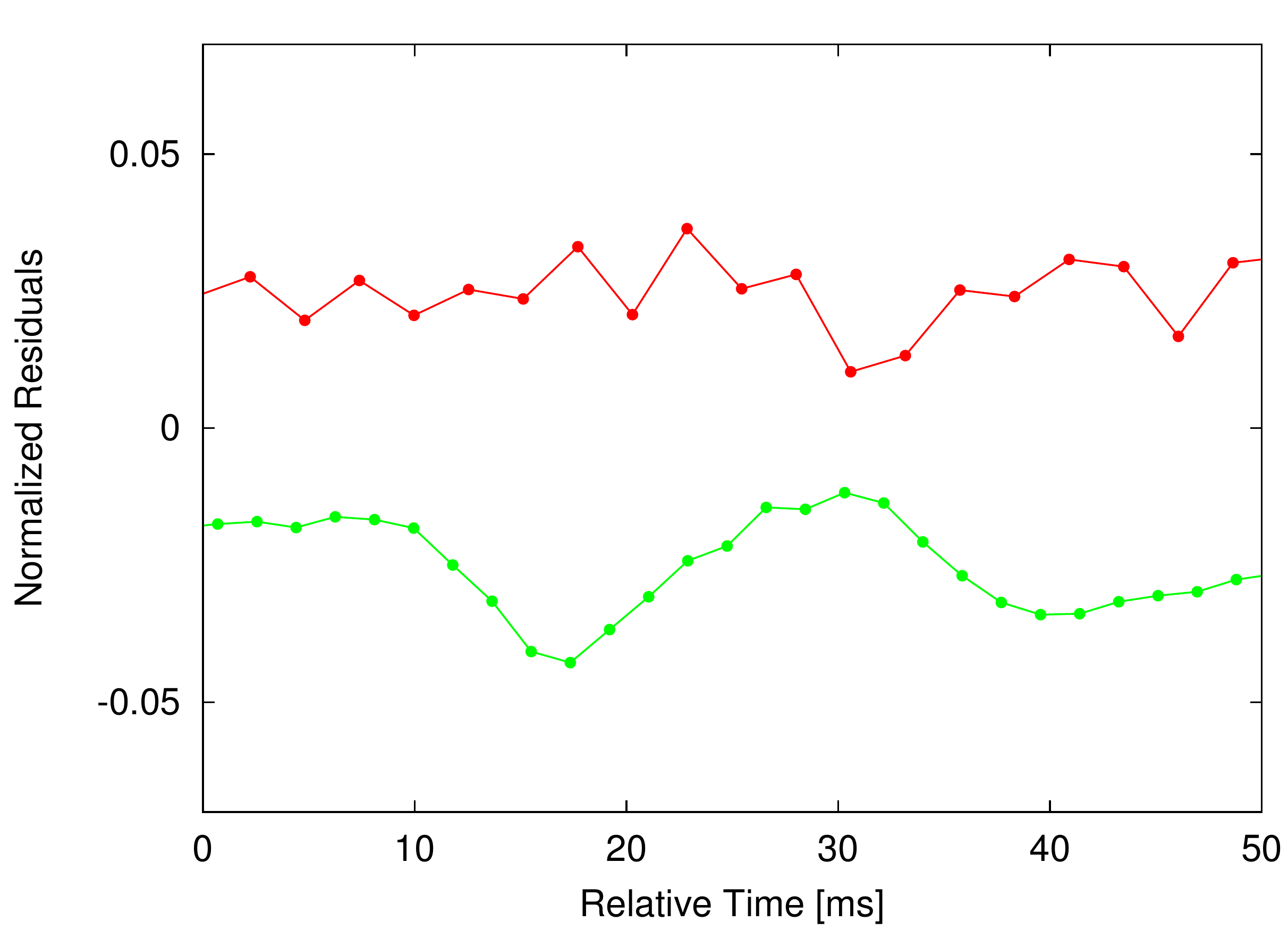}
    \caption{The fit residuals for the central part of the SAO and Devasthal light curves shown in Fig.~\ref{fig:figure1}, top and bottom 
respectively. The residuals have been normalized by the intensity of the unocculted source, and shifted vertically by $\pm0.025$ 
units for clarity. They have also been shifted horizontally into a common
 time window.}
    \label{fig:figure3}
\end{figure}

To investigate further this apparent discrepancy, we have plotted in 
Fig.~\ref{fig:figure3} the fit residuals 
(normalized in units of the light curve intensity) for the two light curves with the best quality, those 
from SAO and from Devasthal, in the central part where the intensity goes from unocculted to 
occulted. It can be noticed that the Devasthal data show a quasi-sinusoidal variation. Scintillation 
comes first to mind as a possible cause, but it would   not be
so regular 
and moreover its amplitude 
should be proportional to the stellar flux. The flux changes from full unocculted intensity at the left 
of Fig.~\ref{fig:figure3}, to zero at the right. Therefore, scintillation can be convincingly excluded as the cause of this 
beating in the residuals for the Devasthal curve. We note that in the case of the SAO data 
scintillation decreases from the right to the left of 
Fig.~\ref{fig:figure3}, 
and indeed one might see that the right 
part of the residuals appears slightly noisier.

We are inclined to conclude that the residuals indicate, at least in the Devasthal case, that the 
UD is not a good model at the level of the accuracy present in the data. The even higher accuracy 
data from SAO however do not show this effect with the same magnitude, and we can speculate 
that the UD (and LD as a consequence) 
model is a better approximation in the infrared part of the visible range.
Another possible reason are time variable changes in the profile,
which we discuss in Sect.~\ref{sect:discuss}.
 The case of the 
TNT data in the ultraviolet is harder to pin down from an analysis of the residuals, since the noise 
was so much higher in this case due to a combination of poorer 
combined throughput
on one hand, and of much higher background on the other. However, we have already remarked 
that the TNT  light curve shown in Fig.~\ref{fig:figure1} seems 
to exhibit a first diffraction fringe which is totally 
unexpected since diffraction effects should have been negligible at this wavelength. 
One possible cause of such a fringe, if real, could be a
compact area (about 20\% of the diameter or less) of enhanced emission
on the photosphere. The jagged ultraviolet brightness profiles
discussed in Sect.~\ref{CAL-DIF} could point in this direction,
although the errors make this far from conclusive.

\subsection{Model-independent Brightness Profiles}\label{CAL-DIF}
LD diameters are  ultimately the values which are employed in standard stellar atmospheric 
models, but 
they are the result of assumed empirical 
conversions which are not directly measured or confirmed by observations. Moreover, the extension 
from the monochromatic values to an effective conversion coefficient valid for a broad bandpass is 
prone to introduce a possible bias. Even more significantly, the whole issue of determining an 
accurate diameter depends on the chosen model brightness distribution which is
assumed to be axisymmetric and constant with time. The  discrepancies which we have 
highlighted in Sect~\ref{sect:ld-discrep} all point to cracks in these assumptions. In the case of Aldebaran however we 
are in the fortunate position to be able to reconstruct the brightness profile directly by 
model-independent methods.

\begin{figure}
	\includegraphics[width=\columnwidth]{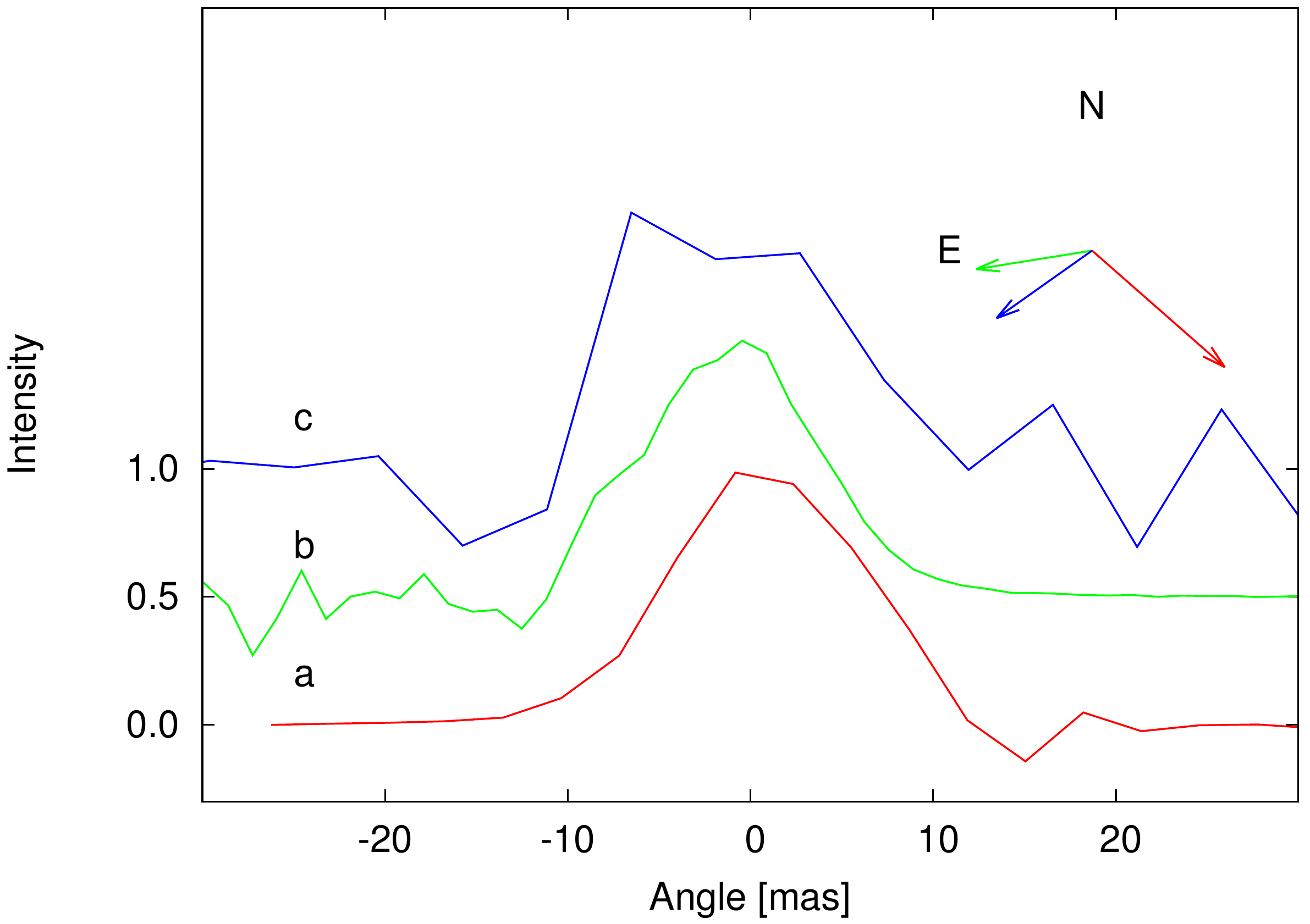}
    \caption{Reconstructed brightness profiles, using differentiation as explained in the text. The curves a, b, c, are for the 
data sets from SAO 6-m, Devasthal 1.3-m and TNT 2.4-m, respectively, with the effective wavelengths listed in Table~\ref{tab:observations}. 
The profiles are shifted by arbitrary amounts in intensity, for the sake of clarity. The arrows display the direction of the 
scan by the lunar limb, projected on the sky and in counter-clockwise
direction from the North they are for b, c, a respectively.
The length of the arrows is inversely proportional to the speed of the scan.}
    \label{fig:figure4}
\end{figure}

Profiles reconstructed by the DIF method are shown in Fig.~\ref{fig:figure4}. By its design, this method is sensitive 
to white noise in the data and this explains why the profiles of the curves from SAO and India are 
noisier on the side where scintillation is affecting the data, the right and left sides of the respective 
profiles. In the case of TNT, the dominant source of noise is not scintillation but rather the shot-noise 
from the very intense background around the Moon in the ultraviolet, and it can be seen that the 
noise in the profile is more evenly distributed. Further, in all three cases a slight dip
towards negative values of the profile is due to presence of small residual diffraction fringe in the light curves: 
this can be seen to the right of the SAO profile, and to the left in the other two cases. It can be 
appreciated how in all three cases the profile does not appear to be completely symmetric. In the 
case of the ultraviolet data from TNT, a relatively flat central part of the profile is present. Limb-brightening as observed in far-ultraviolet solar
images comes to mind, although our SNR is not sufficient to
confirm this hypothesis.

\begin{figure}
	\includegraphics[width=\columnwidth]{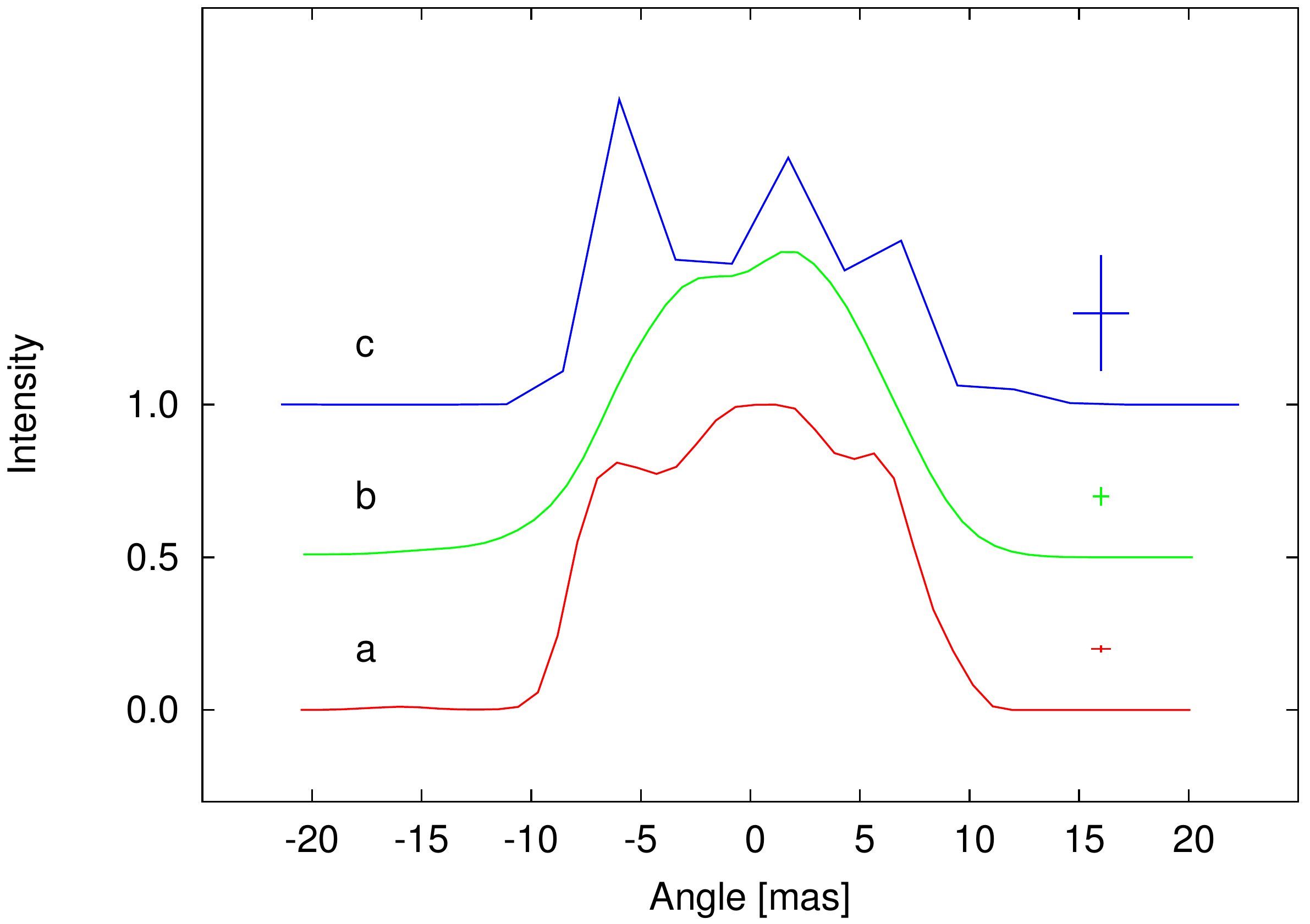}
    \caption{Reconstructed brightness profiles, using the maximum-likelihood CAL method as explained in the text. The 
curves are shifted in intensity, labeled and with the same sky orientation as in Fig.~\ref{fig:figure4}.
The crosses on the right reflect the uncertainties for each profile,
as explained in the text.
}
    \label{fig:figure5}
\end{figure}

We have computed the profiles also by the CAL method, and they are
 shown in Fig.~\ref{fig:figure5}. The 
profiles are consistent with an approximate extent of 20~mas, but show structure which is markedly 
different from the gaussian shape expected for a circularly symmetric uniform disc \citep{1989A&A...226..366R}. 
In the figure we have added error crosses for each profile: the extent in angle is simply the sampling step from Table~\ref{tab:observations},
while the extent in intensity is derived from the SNR of the CAL fit
rescaled by the number of points inside the Aldebaran's disc.
To clarify: the total reconstructed SAO profile extended from
$-50$ to $+40$~mas, and included 46 points with a SNR of 177.8,
or an error of 0.0056 on the normalized intensity.
The points effectively inside the $\pm 10 $~mas extent of the disc
are 22. Assuming that those outside the disc do not contribute
significantly to the noise total, we compute the effective error
in normalized intensity as $46/22\times0.0056=0.012$.

\begin{figure}
	\includegraphics[width=\columnwidth]{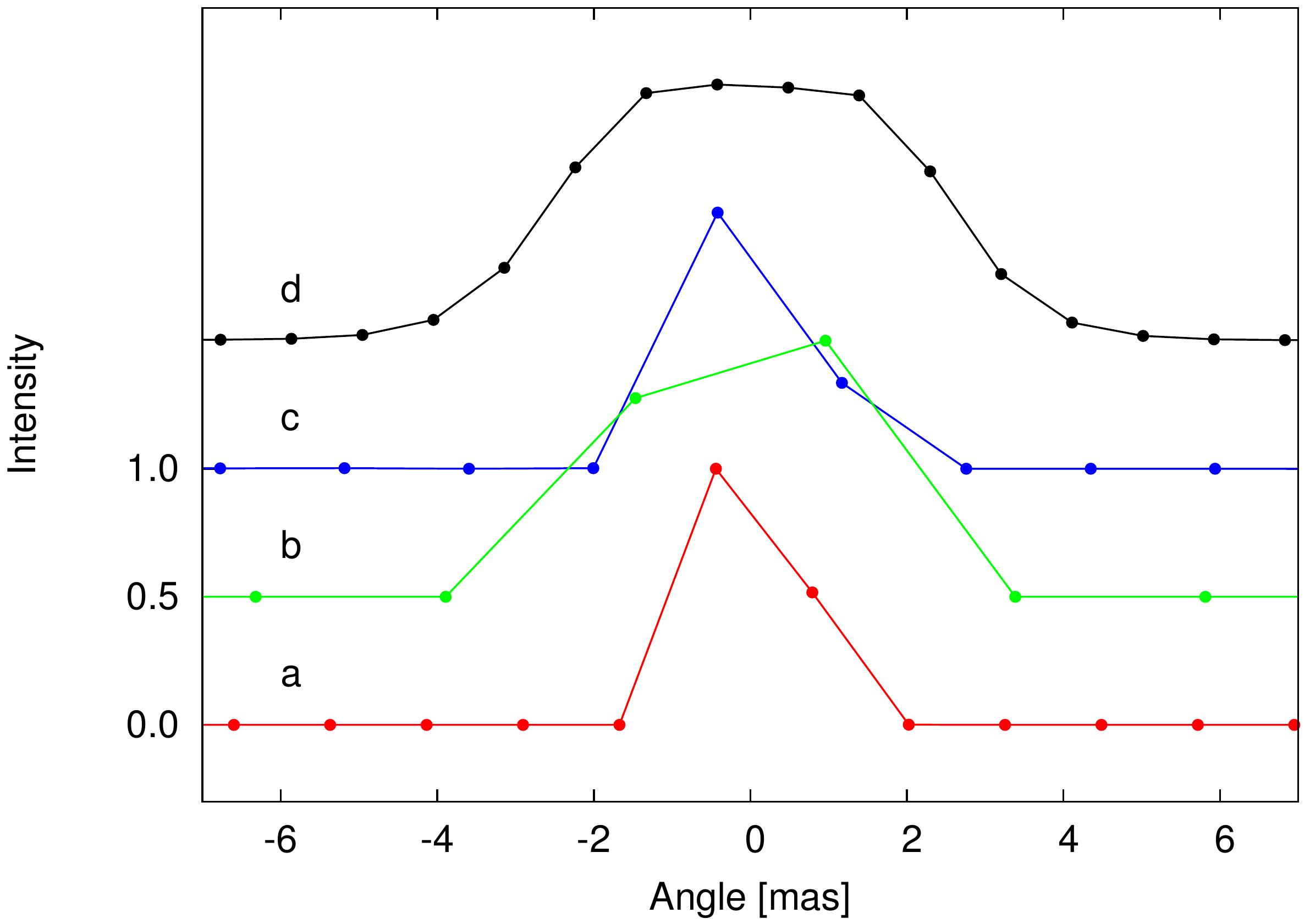}
    \caption{Same as Fig.~\ref{fig:figure5}, for three unresolved sources observed at SAO (a), Devasthal (b) and TNT (c). The curve (d) is for
    a well resolved star observed at SAO. The profiles do
    not display obvious asymmetries.
    The points mark 
the different angular sampling of the data sets. Details in the text.}
    \label{fig:figure6}
\end{figure}

To clarify how these profiles deviate from the expectation, we have analyzed three sources which 
are presumed from empirical considerations to be unresolved, with each of the three telescopes and 
instruments used also for $\alpha$~Tau - albeit in different filters. 
In Fig.~\ref{fig:figure6} we show the CAL profiles 
obtained for 1~Cnc observed at SAO on February 19, 2016; for SAO 94227 observed at Devasthal on 
February 16, 2016; and for HR 4418 observed at TNT on April 18, 2016. 
We have also added a recent LO observation of $\lambda$~Aqr observed 
from SAO on June 25, 2016: this result will be discussed elsewhere,
but it can be seen that the profile is very well resolved and
represents a case not too dissimilar from that of $\alpha$~Tau.
It can be appreciated that, at 
the level of the discrete angular sampling of the three data sets, the 
profiles do not show significant asymmetries, in contrast with those
found for $\alpha$~Tau.

We emphasize that DIF and CAL profiles should be used to investigate the general appearance of the 
brightness profiles only. They are not well suited to measure angular diameters since they are not 
parametric and because the angular scale depends on the adopted limb rate. In our DIF and CAL 
analysis, we adopted the same limb rates as in the LSM method.

\section{Discussion}\label{sect:discuss}
The main common denominator of the results just presented is that while the data can be fitted 
in a first approximation by UD models which can in turn be converted to LD values, in reality all 
three light curves show small deviations which seem to indicate 
the presence of asymmetries 
in the brightness profile. \citetalias{1989A&A...226..366R}
 hinted at a similar possibility from their high 
quality near-IR data, and they 
also mention an unusual scatter in the VLTI long-baseline 
interferometry measurements. Similar findings were reported 
in earlier LO observations of stars with 
very large angular diameters, e.g. asymmetries were hinted in the case of 
$\alpha$~Sco 
\citep{{1957AJ.....62...83E},{1990A&A...230..355R}} - although 
the latter is a supergiant.
For the late-type giant R~Leo,
\citet{1991A&A...249..397D} found from a LO that the UD fit was 
unsatisfactory, and that the brightness profile showed significant
departures from the UD hypothesis.
Studies on other evolved
stars with very large angular diameters have also revealed
significant departures from simple circularly symmetric models,
the best example
being  $\alpha$~Ori for which many different techniques
could be used by several authors 
ranging from adaptive optics to speckle to 
long-baseline interferometry.

One important consideration is that the DIF and CAL profiles that
we obtain appear to be all significantly different from each other.
This is not surprising, when one considers that the data SNR is
quite different in the three cases, and so are the effective
wavelengths. 
In particular, the $u\arcmin$ data from
 TNT represent the shortest wavelength 
 ever used to record a LO of
$\alpha$~Tau, and the photospheric appearance in the
ultraviolet is expected to be considerably different from that in
the red.
Other important factors
are of course the time variability and the different scan
directions. Setting aside for a moment the lower SNR data from TNT,
when one considers e.g. the Devasthal and SAO 
profiles in Fig.~\ref{fig:figure4} it should be noted that
the scan directions were almost orthogonal and that the
two LO events occurred 137$^{\rm d}$ apart. For comparison,
\citet{2015A&A...580A..31H} (who  discovered
a likely exoplanet around $\alpha$~Tau with a 629$^{\rm d}$ period)
attribute residual variations in their radial velocity data 
to rotation modulation of stellar surface features with 
a period of  $\approx 520^{\rm d}$.
These variations could well be related to the photospheric asymmetries 
that we have pointed out, but in this case the time lag
between our LO measurements would be a significant fraction
of the modulation period and hence the comparison of the
SAO and Devasthal data would be problematic.
Obviously, any comparison between the 
present sets of data and earlier ones, such as the LO light curve discussed by \citetalias{1989A&A...226..366R} 
where asymmetries were also suggested, is impossible.

In our LSM analysis discussed in Sect.~\ref{sect:ud-ld} we have
assumed that the lunar limb rate was equal to the predicted value.
We have already stated that the limb rate and the angular
diameter are correlated in a quasi-geometrical optics case
such as that of $\alpha$~Tau, so we tried to prove that 
this assumption is reasonable. We have thus fitted the
SAO data (the set for which we expect the most marked diffraction
effects) leaving the rate as a free parameter. The fitted
best rate was 2.6\% slower than predicted, corresponding to
a limb slope of $2\fdg3$ which would be fully within the norm
for LO events. Significantly however, the quality of the
fit was slightly inferior than for the case with the
predicted rate (normalized $\chi^2$=1.247 instead of 1.206), and
UD angular diameter would then be 18.50~mas instead of 19.12,
also a change in the wrong direction. As expected, in this
case the cross-correlation factor between limb rate
and angular diameter was 0.91. In summary, we think
that our approach of keeping the limb rate fixed to the
predicted value was not detrimental, and in any case
did not affect at all the conclusion on the presence
of photospheric asymmetries.

The most likely physical explanation for such asymmetries in the
brightness profiles would be the presence of surface structures
such as cold spots. Indeed, such spots are expected on
stars like Aldebaran. 
Indirect detection of starspots has been made
possible thanks to
Doppler imaging, and among
the stars included in the
review by \citet{2009A&ARv..17..251S}, about half were
 K giants.
However, starspots are usually more prominent
in fast rotators, and many of the giants for
which starspots are well measured are in
binary systems in which tidal locking accelerates
rotation. An example is the recent extensive study
of XX~Tri by 
\citet{2015A&A...578A.101K}.

In single late-type giants, with periods of the
order of 1-2 years, starspots could not be
detected until recently, but
technological improvements are
starting to reveal  magnetic fields
\citep{2014IAUS..302..350K},
which are
the basis of starspots.
\citet {2009A&A...504..231A, 2015A&A...574A..90A}
have detected sub-Gauss fields in $\beta$~Gem and
$\alpha$~Tau itself, 
and 
\citet{2011A&A...529A.100S} on $\alpha$~Boo.
One additional, indirect support of the starspot
hypothesis is the fact that they are known to have
significantly different contrast in the red and
in the blue: this would help justify further the
differences observed in the brightness profiles.
For example, the
DEV and TNT data are taken at the same time
and along position angles differing by less than $30\degr$,
but the difference in wavelength is dramatic.

As for the magnitude of their effect
on the general brightness profile, the range of starpot sizes
is quite broad. In extreme cases, spots have been
observed to cover about 20\% of the stellar surface,
or down to just fractions of percent at the other extreme.
Their number distribution is also quite varied
since they can be present as single spots or 
in dozens. 
The direct
detection of starspots has recently been demonstrated
 by long-baseline
inteferometry in $\zeta$~And
\citep{2016Natur.533..217R}.
With this work, we show that LO 
of stars with a large angular diameter (above $\approx 10$~mas)
also represent an excellent option
presently available to
measure starspots directly. We note that the Aldebaran occultation
series is ongoing until the end of 2017, and we plan to
observe more events 
from various sites around the northern hemisphere.

As a final remark, Aldebaran is surrounded by a few other stars, at least one of which is suspected of being a physical 
companion. They are all much fainter and with separations of at least tens of arcseconds, and are 
thus undetectable in our light curves and with no influence on our 
findings.

\section{Conclusions}

We have recorded three lunar occultation light curves obtained
first
at the Russian 6-m telescope in the far red, and then 137 days later
 at the Devasthal 1.3-m telescope in the red and at the
2.4-m Thai National Telescope in the ultraviolet. The analysis by
conventional uniform disc (UD) and limb-darkened disc (LD) models
leads to values which are approximately consistent with the
expected LD value of $20.58\pm0.03$\,mas derived by 
\citetalias{1989A&A...226..366R}
from the combination of accurate
occultation and long-baseline interferometry determinations.
However, the measurements do not agree at the level of the
formal errors, and a close inspection of the fit residuals 
showed that the UD (and therefore the LD) approximation may not
be accurate for this K5 giant. This is consistent with earlier
indications of surface asymmetries for Aldebaran as well as for other
late-type giants. 

Analyis by model-independent methods has revealed that the
brightness profile of Aldebaran has significant departures
from spherical symmetry, at least at the milliarcsecond level,
or few percents of its diameter. 
These asymmetries would be well consistent with
cool spots, and lunar occultations provide the
means of detecting such spots directly, if coordinated
observations are performed for the same event from several
sites. We plan to observe more occultations of Aldebaran
in the present series which will last until the end of 2017.

\section*{Acknowledgements}

This work has made use of data obtained at the Thai National Observatory on Doi Inthanon, 
operated by NARIT. We are grateful to Dr.~W.J.~Tango for providing the database of limb-darkening 
corrections,
and to an anonymous referee
for valuable comments and references.
AR acknowledges support from the ESO Scientific Visitor Programme.






\bsp	
\label{lastpage}

\begin{thebibliography}{99}
\bibitem[Auri\`ere et al.(2009)]{2009A&A...504..231A}
Auri\`ere, M., Wade, G.~A., Konstantinova-Antova, R., et al. 2009, A\&A, 504, 231 
\bibitem[Auri{\`e}re et al.(2015)]{2015A&A...574A..90A} 
Auri{\`e}re, M., Konstantinova-Antova, R., Charbonnel, C., et al.\ 2015, A\&A, 574, A90 
\bibitem[Blackwell et al.(1991)]{1991A&A...245..567B} 
Blackwell, D.~E., Lynas-Gray, A.~E., \& Petford, A.~D.\ 1991, A\&A, 245, 567
\bibitem[Bonnell \& Bell(1993)]{1993MNRAS.264..319B} 
Bonnell, J.~T., \& Bell, R.~A.\ 1993, MNRAS, 264, 319
\bibitem[Cayrel de Strobel et al.(1992)]{1992A&AS...95..273C} 
Cayrel de Strobel, G., Hauck, B., Francois, P., et al.\ 1992, A{\&A}S, 95, 273
\bibitem[Davis et al.(2000)]{2000MNRAS.318..387D} 
Davis, J., Tango, W.~J., \& Booth, A.~J.\ 2000, MNRAS, 318, 387
\bibitem[Dhillon et al.(2014)]{2014MNRAS.444.4009D}
Dhillon, V.~S., Marsh, T.~R., Atkinson, D.~C., et al.\ 2014, MNRAS, 444, 4009
\bibitem[di Giacomo et al.(1991)]{1991A&A...249..397D} 
di Giacomo, A., Lisi, F., Calamai, G., \& Richichi, A.\ 1991, A{\&A}, 249, 397
\bibitem[Evans(1957)]{1957AJ.....62...83E} Evans, D.~S.\ 1957, AJ, 62, 83
\bibitem[Hatzes et al.(2015)]{2015A&A...580A..31H} Hatzes, 
A.~P., Cochran, W.~D., Endl, M., et al.\ 2015, A{\&A}, 580, A31
\bibitem[Jofr{\'e} et al.(2014)]{2014A&A...564A.133J} 
Jofr{\'e}, P., Heiter, U., Soubiran, C., et al.\ 2014, A{\&A}, 564, A133
\bibitem[Korhonen(2014)]{2014IAUS..302..350K} Korhonen, 
H.\ 2014, Magnetic Fields throughout Stellar Evolution, 302, 350
\bibitem[K{\"u}nstler et al.(2015)]{2015A&A...578A.101K} 
K{\"u}nstler, A., Carroll, T.~A., \& Strassmeier, K.~G.\ 2015, A{\&A}, 578, A101
\bibitem[Richichi(1989)]{1989A&A...226..366R} 
Richichi, A.\ 1989, A{\&A}, 226, 366
\bibitem[Richichi \& Lisi(1990)]{1990A&A...230..355R}
 Richichi, A., \& Lisi, F.\ 1990, A{\&A}, 230, 355
\bibitem[Richichi et al.(1992)]{1992A&A...265..535R} 
Richichi, A., di Giacomo, A., Lisi, F., \& Calamai, G.\ 1992, A{\&A}, 265, 535
\bibitem[Richichi \& Roccatagliata(2005)]{2005A&A...433..305R} 
Richichi, A., \& Roccatagliata, V.\ 2005, A{\&A}, 433, 305 (RR05)
\bibitem[Richichi et al.(2008)]{2008A&A...489.1399R} Richichi, A., Fors, O., Mason, E., Stegmeier, J., 
\& Chandrasekhar, T.\ 2008, A{\&A}, 489, 1399
\bibitem[Richichi et al.(2014)]{2014AJ....148..100R} Richichi, A., Irawati, P., Soonthornthum, B., 
Dhillon, V.~S., \& Marsh, T.~R.\ 2014, AJ, 148, 100
\bibitem[Richichi et al.(2016)]{2016AJ....151...10R} Richichi, A., Tasuya, O., Irawati, P., et al.\ 2016, 
AJ, 151, 10
\bibitem[Roettenbacher et al.(2016)]{2016Natur.533..217R} Roettenbacher, R.~M., Monnier, J.~D., Korhonen, H., et al.\ 2016, Nature, 533, 217 
\bibitem[Sennhauser \& Berdyugina(2011)]{2011A&A...529A.100S} Sennhauser, C., \& Berdyugina, S.~V.\ 2011, A{\&A}, 529, A100
\bibitem[Strassmeier(2009)]{2009A&ARv..17..251S} Strassmeier, K.~G.\ 2009, A{\&A}Rv, 17, 251

\end{thebibliography}
\end{document}